\documentclass[twocolumn,english,pra,showpacs]{revtex4-1}
\usepackage[T1]{fontenc}
\usepackage[latin9]{inputenc}
\usepackage{amsmath}
\usepackage{amssymb}
\usepackage{graphicx}

\makeatletter

\providecommand{\tabularnewline}{\\}

\@ifundefined{textcolor}{}
{%
 \definecolor{BLACK}{gray}{0}
 \definecolor{WHITE}{gray}{1}
 \definecolor{RED}{rgb}{1,0,0}
 \definecolor{GREEN}{rgb}{0,1,0}
 \definecolor{BLUE}{rgb}{0,0,1}
 \definecolor{CYAN}{cmyk}{1,0,0,0}
 \definecolor{MAGENTA}{cmyk}{0,1,0,0}
 \definecolor{YELLOW}{cmyk}{0,0,1,0}
}

\usepackage{array}
\usepackage{multirow}

\usepackage{babel}

\makeatother

\usepackage{babel}
\begin{document}

\title{Amplification limit of weak measurements: a variational approach}

\author{Shengshi Pang$^{1}$}

\email{shengshp@usc.edu}

\author{Todd A. Brun$^{1}$}

\email{tbrun@usc.edu}

\author{Shengjun Wu$^{2,3}$}

\email{sjwu@nju.edu.cn}

\author{Zeng-Bing Chen$^{2}$}

\email{zbchen@ustc.edu.cn}

\affiliation{$^{1}$Department of Electrical Engineering, University of Southern
California, Los Angeles, California 90089, USA}

\affiliation{$^{2}$Hefei National Laboratory for Physical Sciences at Microscale
and Department of Modern Physics, University of Science and Technology
of China, Hefei, Anhui 230026, China}

\affiliation{$^{3}$Kuang Yaming Honors School, Nanjing University, Nanjing 210093,
China}
\begin{abstract}
Post-selected weak measurement has been widely used in experiments
to observe weak effects in various physical systems. However, it is
still unclear how large the amplification ability of a weak measurement
can be and what determines the limit of this ability, which is fundamental
to understanding and applying weak measurements. The limitation of
the conventional weak value formalism for this problem is the divergence
of weak values when the pre- and post-selections are nearly orthogonal.
In this paper, we study this problem by a variational approach for
a general Hamiltonian $H_{\mathrm{int}}=gA\otimes\Omega\delta(t-t_{0}),\, g\ll1$.
We derive a general asymptotic solution, and show that the amplification
limit is essentially independent of $g$, and determined only by the
initial state of the detector and the number of distinct eigenvalues
of $A$ or $\Omega$. An example of spin-$\frac{1}{2}$ particles
with a pair of Stern-Gerlach devices is given to illustrate the results.
The limiting case of continuous variable systems is also investigated
to demonstrate the influence of system dimension on the amplification
limit. 
\end{abstract}


\maketitle
\global\long\def\i{\mathrm{i}}
 \global\long\def\e{\mathrm{e}}
 \global\long\def\re{\mathrm{Re}}
 \global\long\def\im{\mathrm{Im}}
 \global\long\def\d{\mathrm{d}}
 \global\long\def\tr{\mathrm{tr}}
 \global\long\def\ds{d_{\mathrm{s}}}
 \global\long\def\bu{\boldsymbol{\mu}}
 \global\long\def\wx{\widetilde{\Xi}}
 \global\long\def\dd{d_{\mathrm{D}}}
 \global\long\def\ra{r_{A}}
 \global\long\def\ro{r_{\Omega}}
 \global\long\def\hd{\mathcal{H}_{\mathrm{D}}}

\section{Introduction}

The von Neumann projective measurement model is well known theory
for standard quantum measurements, in which the readings of a measurement
are the eigenvalues of an observable, and the system is projected
into the corresponding eigenstate of the observable. To realize such
an ideal measurement, the spread width of the detector's wave functions
must be sufficiently narrow, or the interaction between the system
and the detector must be sufficiently strong, so that the detector's
final states---translated by different eigenvalues of the observable---can
be distinguished with high probability.

In contrast to von Neumann measurements, \emph{weak measurements}
(coined by Aharonov, Albert, and Vaidman in 1988 \cite{AAV}) exploit
the opposite conditions: the initial detector state has a very wide
spread, or the measurement strength is ultra-weak. Such a weak measurement
makes the detector's final states, translated by different eigenvalues
of the system observable, significantly overlap with each other. And
a further step of this protocol, postselection, superposes these states.
Interference between them can dramatically change the original state
of the detector (not only by a translation). A remarkable effect induced
by this interference is that the output from the postselected detector
can be much larger than the eigenvalue spectrum of the system observable,
due to the coherence in the superposed state of the detector canceling
the major part of the original detector wave function \cite{prd}.

This striking difference from standard von Neumann measurements makes
weak measurement particularly useful to amplify small physical quantities.
Experiments have successfully realized the amplification of many different
physical effects by postselected weak measurements, including the
spin Hall effect of light \cite{science spin hall,spin hall 2}, optical
beam deflection \cite{beam-deflection1,beam-deflection2}, optical
frequency shift \cite{optical frequency}, optical phase shift \cite{phase amplification,phase amplification2},
temperature shift \cite{temperature shift}, longitudinal velocities
\cite{longitudinal velocities}, etc. More experimental protocols
have been proposed \cite{proposal-chargesensing,proposal-electron spin,proposal-phaseshift,proposal-Tomography of Many-Body Weak Value,proposal-wu-marek,proposal-time delay of light,proposal-entangled1,proposal-entangled2,proposal-Justin-recycled photons}.
Moreover, weak measurements have been realized on various physical
systems besides optical systems, including SQUIDs \cite{cqed,cqed2 violation of Leggett-Garg inequality}
and NMR \cite{nmr}, among others.

Despite the considerable existing research on the theory of weak measurements,
and their increasing application in experiments, a fundamental problem
is still unclear: what is the ultimate limit of amplification in a
postselected weak measurement? Usually, weak measurements are studied
in the weak value formalism (see \cite{liyang} for a general framework,
and \cite{physics report,justin rmp} for reviews). However, in the
weak value formalism, the amplification of a weak measurement can
be infinitely large if the inner product between the pre- and postselections
of the system is sufficiently small. From a practical view, this is
obviously impossible. The root of this contradiction is that the weak
value formalism is valid only when the amplification is small, since
it is a first-order approximation theory that works only when the
response of the detector is linear in the signal. When the amplification
is too strong, the response of the detector to the signal becomes
nonlinear, so the weak value formalism breaks down. It cannot give
a valid result for the limit of amplification in a postselected weak
measurement.

The significance of this problem is manifold: (1) it determines clearly
to what extent postselected weak measurements can amplify small signals
in practical experiments, thus revealing the limits of the quantum
advantage in this task; (2) it can show what determines the amplification
limit of a weak measurement, thus providing guidance for designing
experiments; (3) it goes beyond the limitation of the weak value formalism,
so its result will be an important supplement to the current knowledge
of weak measurement. Given the wide application of weak measurements
in many different branches of physics, the solution of this problem
will be broadly useful to the physics community.

Despite the importance of this problem, few results have been known
to date, and a complete solution is still missing. Numerical upper
bounds were observed for some cases in \cite{numerical bound,liyang}.
Certain special cases were studied with specific assumptions on the
observable $A$ (e.g., $A^{2}=I$ in \cite{a2=00003D00003DI1,a2=00003D00003DI2}),
or on the detector states (e.g., a qubit system with a Gaussian detector
in \cite{zhu}). Orthogonal and asymptotically orthogonal pre- and
postselections were considered in \cite{my paper}. An optimal detector
for a given experimental setup was provided in \cite{optimal probe}.
In a more recent paper \cite{antonoio}, a refined weak value method
was attempted; but it is still not sufficient to give a rigorous solution
without including higher order weak values---particularly when the
dimension of the system or the detector is high---since higher order
weak values can dominate over the lowest order weak value. (This has
been verified by weak measurements with OAM pointer states \cite{oam}.)

In this paper, we fill this gap by exploiting a variational approach
to study the problem in a rigorous way. We derive a general analytical
solution for a weak coupling Hamiltonian $H_{\mathrm{int}}=gA\otimes\Omega\delta(t-t_{0}),\, g\ll1$,
and reveal the surprising property that the solution is independent
of the coupling strength $g$ when $g\ll1$, depending only on the
initial state of the detector and the dimension of the system or the
detector, whichever is less (if $A$ and $\Omega$ are nondegenerate).
This is in marked contrast to the weak value formalism. For degenerate
$A$ or $\Omega$, the degeneracy will decrease the amplification
limit, which depends on the number of distinct eigenvalues. The results
are illustrated in detail by an example of spin-$\frac{1}{2}$ particles
passing through a pair of Stern-Gerlach devices. We also consider
continuous variable systems as a limiting example, to show how the
dimension of the system can significantly influence the amplification
limit, which also is missed by the weak value formalism.

\section{Preliminary: the weak value formalism}

We start by revisiting the weak value formalism. Suppose the initial
state of the system is $|\Psi_{i}\rangle$, the postselected state
is $|\Psi_{f}\rangle$, the initial state of the detector is $|\Upsilon\rangle$,
and the interaction Hamiltonian between the system and the detector
is 
\begin{equation}
H_{\mathrm{int}}=gA\otimes\Omega\delta(t-t_{0}).\label{eq:33-1}
\end{equation}
Let $\hbar=1$. The final state of the detector after the interaction
followed by the postselection is $|\Upsilon_{f}\rangle=\langle\Psi_{f}|\exp(-\i gA\otimes\Omega)|\Psi_{i}\rangle|\Upsilon\rangle,$
where $|\Upsilon_{f}\rangle$ is unnormalized. If we measure an observable
$M$ on the final detector state, the detector will have a shift of
$\frac{\langle\Upsilon_{f}|M|\Upsilon_{f}\rangle}{\langle\Upsilon_{f}|\Upsilon_{f}\rangle}-\langle\Upsilon|M|\Upsilon\rangle$
in the expected value of $M$.

When $g$ is sufficiently small, the final state of the detector is
approximately $\exp(-\i gA_{w}\Omega)|\Upsilon\rangle$, where 
\begin{equation}
A_{w}=\frac{\langle\Psi_{f}|A|\Psi_{i}\rangle}{\langle\Psi_{f}|\Psi_{i}\rangle},
\end{equation}
is called the \emph{weak value}. It can be derived that the average
output of the detector is 
\begin{equation}
\begin{aligned} & g\im A_{w}(\langle\Upsilon|\{\Omega,M\}|\Upsilon\rangle-2\langle\Upsilon|\Omega|\Upsilon\rangle\langle\Upsilon|M|\Upsilon\rangle)\\
 & +\i g\re A_{w}\langle\Upsilon|[\Omega,M]|\Upsilon\rangle.
\end{aligned}
\label{eq:17}
\end{equation}
The second term of Eq.~(\ref{eq:17}) is real because the average
of a commutator must be imaginary.

One can see from (\ref{eq:17}) that the average output of a weak
measurement approximately amplifies $g$ by the real or imaginary
part of the weak value. This is the basis of all weak measurement
amplification protocols. It is worth mentioning that Ref. \cite{josza}
demonstrated the roles of the real and imaginary parts of the weak
value in the position or momentum shift of the detector in a postselected
weak measurement. Eq.~(\ref{eq:17}) is a generalization of those
results to an arbitrary observable $M$ on the detector, and when
$[\Omega,M]=0\,{\rm or}\,\i$, (\ref{eq:17}) reduces to the results
of \cite{josza}.

An obvious shortcoming of the weak value formalism is that when $\langle\Psi_{f}|\Psi_{i}\rangle\rightarrow0$,
$A_{w}\rightarrow\infty$, implying that the output of a weak measurement
could be infinite, which is impossible in practice. This issue is
rooted in the first order approximation in deriving the weak value
formalism, so it is not a proper tool to study the amplification limit
of a weak measurement.

\section{Amplification limit: a variational approach}

In the following, we shall show a variational approach to this problem
that can avoid the divergence of the weak value formalism and give
a valid result for the amplification limit.

Let's define a shift operator $\Delta M=M-\langle\Upsilon|M|\Upsilon\rangle$.
The average shift of the detector is 
\begin{equation}
\langle\Delta M\rangle=\frac{\langle\Upsilon_{f}|\Delta M|\Upsilon_{f}\rangle}{\langle\Upsilon_{f}|\Upsilon_{f}\rangle}.
\end{equation}
When we choose different pre- and postselections of the system, the
detector will give different outputs at the end of the measurement.
Intuitively, the shift of the detector should be bounded. Our goal
is to find the maximum $\langle\Delta M\rangle$ over all possible
pre- and postselections. This maximum is the \emph{amplification limit}.

When $\langle\Delta M\rangle$ attains an extremal value $\langle\Delta M\rangle_{\mathrm{e}}$,
its variation with respect to $|\Upsilon_{f}\rangle$ is zero: 
\begin{equation}
\begin{aligned}\delta\langle\Delta M\rangle_{\e} & =\frac{1}{\langle\Upsilon_{f}|\Upsilon_{f}\rangle}((\delta\langle\Upsilon{}_{f}|)(\Delta M|\Upsilon_{f}\rangle-|\Upsilon_{f}\rangle\langle\Delta M\rangle_{\e})\\
 & +(\langle\Upsilon_{f}|\Delta M-\langle\Delta M\rangle_{\e}\langle\Upsilon_{f}|)(\delta|\Upsilon_{f}\rangle))=0,
\end{aligned}
\end{equation}
according to the variational principle. Since $|\Upsilon_{f}\rangle$
is determined by $|\Psi_{i}\rangle$ and $|\Psi_{f}\rangle$, the
variation of $|\Upsilon_{f}\rangle$ can be expressed in terms of
the variations of $|\Psi_{i}\rangle$ and $|\Psi_{f}\rangle$: 
\begin{equation}
\begin{aligned}\delta|\Upsilon_{f}\rangle & =(\delta\langle\Psi_{f}|)\exp(-\i gA\otimes\Omega)|\Psi_{i}\rangle|\Upsilon\rangle\\
 & +\langle\Psi_{f}|\exp(-\i gA\otimes\Omega)(\delta|\Psi_{i}\rangle)|\Upsilon\rangle.
\end{aligned}
\end{equation}
Thus, the variation of $\langle\Delta M\rangle_{\e}$ becomes 
\begin{equation}
\begin{alignedat}{1} & \delta\langle\Delta M\rangle_{\e}\\
 & =\langle\Upsilon|\langle\Psi_{i}|\exp(\i gA\otimes\Omega)(\Delta M|\Upsilon_{f}\rangle-|\Upsilon_{f}\rangle\langle\Delta M\rangle_{\e})(\delta|\Psi_{f}\rangle)\\
 & +(\delta\langle\Psi_{i}|)\langle\Upsilon|\exp(\i gA\otimes\Omega)|\Psi_{f}\rangle(\Delta M|\Upsilon_{f}\rangle-|\Upsilon_{f}\rangle\langle\Delta M\rangle_{\e})\\
 & +\mathrm{c.c.}=0.
\end{alignedat}
\label{eq:8}
\end{equation}
Note that the variations $\delta|\Psi_{i}\rangle$ and $\delta|\Psi_{f}\rangle$
are arbitrary and independent, it follows from (\ref{eq:8}) that
\begin{equation}
\begin{alignedat}{1}\langle\Upsilon|\exp(\i gA\otimes\Omega)|\Psi_{f}\rangle(\Delta M-\langle\Delta M\rangle_{\e})|\Upsilon_{f}\rangle & =0,\\
\langle\Upsilon|\langle\Psi_{i}|\exp(\i gA\otimes\Omega)(\Delta M-\langle\Delta M\rangle_{\e})|\Upsilon_{f}\rangle & =0.
\end{alignedat}
\label{eq:7}
\end{equation}

It is crucial to observe that the left sides of the two equations
in (\ref{eq:7}) are vectors in the system Hilbert space, so their
amplitudes in the eigenbasis of $A$ must all be zero. Suppose the
observable $A$ has $\ra$ eigenvalues $a_{1},\cdots,a_{\ds}$ with
eigenstates $|a_{1}\rangle,\cdots,|a_{\ds}\rangle$. If $|\Psi_{i}\rangle=\sum_{k}\alpha_{k}|a_{k}\rangle,\,|\Psi_{f}\rangle=\sum_{k}\beta_{k}|a_{k}\rangle$,
the two equations of (\ref{eq:7}) give 
\begin{equation}
\begin{aligned}\sum_{k=1}^{\ds}\alpha_{k}|a_{k}\rangle\langle\Upsilon|\exp(\i ga_{k}\Omega)(\Delta M-\langle\Delta M\rangle_{\e})|\Upsilon_{f}\rangle & =0,\\
\sum_{k=1}^{\ds}\beta_{k}^{*}\langle a_{k}|\langle\Upsilon|\exp(\i ga_{k}\Omega)(\Delta M-\langle\Delta M\rangle_{\e})|\Upsilon_{f}\rangle & =0.
\end{aligned}
\end{equation}
Since either $|\Psi_{i}\rangle$ or $|\Psi_{f}\rangle$ can be arbitrary,
we can choose all $\alpha_{k}\neq0$ or all $\beta_{k}\neq0$. Then
the above equation shows 
\begin{equation}
\langle\Upsilon|\exp(\i ga_{i}\Omega)(\Delta M-\langle\Delta M\rangle_{\e})|\Upsilon_{f}\rangle=0
\end{equation}
for all $i=1,\cdots,\ra$.

Let $\Xi_{g}$ be the matrix $\left[\e^{-\i ga_{1}\Omega}|\Upsilon\rangle|\cdots|\e^{-\i ga_{\ds}\Omega}|\Upsilon\rangle\right]$.
We can write $|\Upsilon_{f}\rangle$ as a linear combination of $\exp(-\i ga_{k}\Omega)|\Upsilon\rangle,\, k=1,\cdots,\ds$,
i.e., $|\Upsilon_{f}\rangle=\Xi_{g}\bu$, $\bu=(\mu_{1},\cdots,\mu_{\ds})^{T}$,
and $\mu_{k}=\beta_{k}^{*}\alpha_{k}$. Then 
\begin{equation}
(\Xi_{g}^{\dagger}\Delta M\Xi_{g}-\Xi_{g}^{\dagger}\Xi_{g}\langle\Delta M\rangle_{\e})\bu=0,\label{eq:20}
\end{equation}
Eq.~(\ref{eq:20}) is a homogeneous linear equation with respect
to $\bu$, so the necessary and sufficient condition for the existence
of nonzero $\boldsymbol{\mu}$ is 
\begin{equation}
\det(\Xi_{g}^{\dagger}\Delta M\Xi_{g}-\Xi_{g}^{\dagger}\Xi_{g}\langle\Delta M\rangle_{\e})=0,\label{eq:5}
\end{equation}
which is the major equation that $\langle\Delta M\rangle_{\e}$ must
satisfy.

Eq.~(\ref{eq:5}) implies that an extremum of $\langle\Delta M\rangle$
must be an eigenvalue of $(\Xi_{g}^{\dagger}\Xi_{g})^{-\frac{1}{2}}\Xi_{g}^{\dagger}\Delta M\Xi_{g}(\Xi_{g}^{\dagger}\Xi_{g})^{-\frac{1}{2}}$.
Therefore, the largest $\langle\Delta M\rangle_{\e}$ is its largest
eigenvalue: 
\begin{equation}
|\langle\Delta M\rangle|_{\max}=|\lambda((\Xi_{g}^{\dagger}\Xi_{g})^{-\frac{1}{2}}\Xi_{g}^{\dagger}\Delta M\Xi_{g}(\Xi_{g}^{\dagger}\Xi_{g})^{-\frac{1}{2}})|_{\max}.\label{eq:1}
\end{equation}
Note that above we have assumed $\Xi_{g}$ to be full rank, so that
$(\Xi_{g}^{\dagger}\Xi_{g})^{\frac{1}{2}}$ is invertible. If the
eigenvalues of $A$ are degenerate, or $\ds>\dd$, the rank of $\Xi_{g}$
will be less than $\ds$; in that case, one needs to pick out a maximal
linearly independent subset from $\exp(-\i ga_{i}\Omega)|\Upsilon\rangle,\, i=1,\cdots,\ds$
to construct the matrix $\Xi_{g}$.

A formal asymptotic solution for the amplification limit can be obtained
from (\ref{eq:1}) by Gelfand's theorem \cite{Gelfand's theorem}
which connects the spectral radius of a matrix to its (arbitrary)
norm. If we choose the norm $\left\Vert \cdot\right\Vert $ to be
the trace norm, then 
\begin{equation}
|\langle\Delta M\rangle|_{\max}=\lim_{n\rightarrow\infty}(\tr(\Xi_{g}^{\dagger}\Delta M\Xi_{g}(\Xi_{g}^{\dagger}\Xi_{g})^{-1})^{n})^{\frac{1}{n}}.\label{eq:24}
\end{equation}
Usually, one can choose a finite $k$ to derive an approximate solution
to $|\langle\Delta M\rangle|_{\max}$ from (\ref{eq:24}), and large
$k$ will give higher precision to the approximation.

In a weak measurement, the coupling strength $g$ is usually very
small. This can lead to a simplified and $g$-independent form of
(\ref{eq:5}), which is helpful in finding the essential factors that
determine the amplification limit.

The key is to prove that when $g\ll1$, the support of $\Xi_{g}$,
i.e., the subspace spanned by all translated detector states $\exp(-\i ga_{k}\Omega)|\Upsilon\rangle,\, k=1,\cdots,\ds$,
is $g$-independent, and can be spanned approximately by $|\Upsilon\rangle,\Omega|\Upsilon\rangle,\cdots,\Omega^{\ra-1}|\Upsilon\rangle$
(unorthonormalized), where $\ra$ is the number of different eigenvalues
of $A$. The proof is given in Appendix A. If we define $\wx$ to
be the matrix $\left[|\Upsilon\rangle|\,\Omega|\Upsilon\rangle\,|\,\cdots\,|\,\Omega^{\ra-1}|\Upsilon\rangle\right]$,
the equation for $\langle\Delta M\rangle_{\e}$ (\ref{eq:5}) can
be simplified to 
\begin{equation}
\det(\wx^{\dagger}\Delta M\wx-\wx^{\dagger}\wx\langle\Delta M\rangle_{\e})=0.\label{eq:6}
\end{equation}

If $\wx$ is full rank, Eq.~(\ref{eq:6}) is equivalent to the eigenvalue
equation for $(\wx^{\dagger}\wx)^{-\frac{1}{2}}\wx^{\dagger}\Delta M\wx(\wx^{\dagger}\wx)^{-\frac{1}{2}}$,
so 
\begin{equation}
|\langle\Delta M\rangle|_{\max}=|\lambda((\wx^{\dagger}\wx)^{-\frac{1}{2}}\wx^{\dagger}\Delta M\wx(\wx^{\dagger}\wx)^{-\frac{1}{2}})|_{\max}.\label{eq:13}
\end{equation}
If the rank of $\wx$ is less than $\ra$, one just needs to pick
out a maximal linearly independent set from $|\Upsilon\rangle,\Omega|\Upsilon\rangle,\cdots,\Omega^{\ra-1}|\Upsilon\rangle$
to reconstruct $\wx$. So (\ref{eq:13}) will still hold.

Eq.~(\ref{eq:13}) is a general solution for the amplification limit
$|\langle\Delta M\rangle|_{\max}$. An explicit asymptotic solution
can also be derived from (\ref{eq:13}) by Gelfand's theorem \cite{Gelfand's theorem},
which connects the spectral radius of a matrix to its (arbitrary)
norm. If we choose the norm $\left\Vert \cdot\right\Vert $ in \cite{Gelfand's theorem}
to be the trace norm, then 
\begin{equation}
|\langle\Delta M\rangle|_{\max}=\lim_{n\rightarrow\infty}(\tr(\wx^{\dagger}\Delta M\wx(\wx^{\dagger}\wx)^{-1})^{n})^{\frac{1}{n}}.\label{eq:4}
\end{equation}

We see from (\ref{eq:13}) and (\ref{eq:4}) that the amplification
limit $|\langle\Delta M\rangle|_{\max}$ is independent of $g$, and
determined only by the subspace spanned by $|\Upsilon\rangle,\Omega|\Upsilon\rangle,\cdots,\Omega^{\ra-1}|\Upsilon\rangle$.
Denote this subspace as $\hd$. Usually, when $\ra$ increases, $\hd$
will become larger, so $|\langle\Delta M\rangle|_{\max}$ will increase
as well. But if $\ra>\ro$, where $\ro$ is the number of distinct
eigenvalues of $\Omega$, then this subspace will be $\ro$ dimensional
at most, and the increase of $\ra$ will no longer affect $|\langle\Delta M\rangle|_{\max}$.
In addition, the size of the support of $|\Upsilon\rangle$ on the
eigenbasis of $\Omega$ also influences this subspace. For example,
if $|\Upsilon\rangle$ is an eigenstate of $\Omega$, then $\hd$
is one dimensional, whatever $\ra$ and $\ro$ are. A detailed analysis
is given in Appendix B.

\section{Examples}

\subsection{Spin-$\frac{1}{2}$ particles}

To illustrate our result, we consider an example of spin-$\frac{1}{2}$
particles with a pair of Stern-Gerlach devices \cite{AAV}.

When a beam of spin-$\frac{1}{2}$ particles with the same spin direction
moving in the $x$ direction pass through a Stern-Gerlach device which
has a nonuniform external magnetic field in the $z$ direction, it
will be coupled to the magnetic field due to the interaction 
\begin{equation}
H_{I}=-\mu\frac{\partial B_{z}}{\partial z}z\sigma_{z}\delta(x-x_{0}),\label{eq:3}
\end{equation}
where $\mu$ is the magnetic moment of a single particle, and $\delta(x-x_{0})$
means that the support of the magnetic field is very narrow so that
the duration of the interaction is extremely short. Immediately after
the first Stern-Gerlach device, let the beam of particles pass through
a second Stern-Gerlach device, where the magnetic field is in the
$y$ direction. Then the particles will split into two beams with
spins pointing to the $\pm y$ directions respectively. If we keep
track of one of the two beams, say the beam with spins $+y$, then
we actually postselect the particles in the state of spin $+y$. If
the gradient of the magnetic field in the first Stern-Gerlach device
is sufficiently small, i.e. $|\frac{\partial B_{z}}{\partial z}|\ll1$,
and the initial direction of the spins is properly chosen to be close
to the $-y$ direction, then the $+y$ beam will have a large displacement
in the $z$ direction due to the amplification effect of the postselected
weak measurement.

Now, we can apply the results derived above to find the largest possible
displacement of the postselected beam over all directions of the magnetic
field in the second Stern-Gerlach device. In this example, $g=-\mu\frac{\partial B_{z}}{\partial z},\,\Omega=z,\, M=z,\, p_{z}$.
Suppose the initial spatial wave function of the particles in the
$z$ direction is $\Upsilon(z)$ and is symmetric about its center
for simplicity. Since $\ds=2$ for spin-$\frac{1}{2}$ particles,
the final spatial wave function of the particles can be spanned by
$\Upsilon(z)$ and $z\Upsilon(z)$ (unnormalized). If the position
or the momentum of the particle is measured in the $z$ direction
after the second Stern-Gerlach device, then from Eq.~(\ref{eq:13})
one can obtain (see Appendix C) 
\begin{equation}
\begin{aligned}|\langle\Delta z\rangle|_{\max} & =\sqrt{\langle\Delta z^{2}\rangle_{\Upsilon}},\\
|\langle\Delta p_{z}\rangle|_{\max} & =\frac{1}{2}\sqrt{\frac{1+\langle\{z,\Delta p_{z}\}\rangle_{\Upsilon}^{2}}{\langle\Delta z^{2}\rangle_{\Upsilon}}},
\end{aligned}
\label{eq:9}
\end{equation}
where $\langle\Upsilon|\cdot|\Upsilon\rangle$ is denoted $\langle\cdot\rangle_{\Upsilon}$
for brevity. From (\ref{eq:9}), one can see that neither $|\langle\Delta z\rangle|_{\max}$
nor $|\langle\Delta p_{z}\rangle|_{\max}$ depend on the magnetic
moment $\mu$ or the external field $B_{z}$, and they are determined
only by the initial wave function of the particle, which confirms
the previous result.

We applied (\ref{eq:9}) to three typical spatial wave functions for
the particles: the Gaussian state, the Lorentzian state and the exponential
state; the results are summarized in Table I. In addition, the results
for the exponential state are plotted in Fig. I for different weak
values. It can be seen that upper bounds always exist whatever $g$
is, and the upper bounds are nearly the same when $g\ll1$, which
verifies the previous result. 
\begin{table}
\begin{tabular}{cccc}
\hline 
Detector state  & Detector wave function  & \multirow{1}{*}{$|\langle\Delta z\rangle|_{\max}$}  & $|\langle\Delta p_{z}\rangle|_{\max}$\tabularnewline
\hline 
\hline 
Gaussian  & $((2\pi)^{\frac{1}{2}}K)^{-\frac{1}{2}}\exp(-\frac{z^{2}}{4K^{2}})$  & $K$  & $\frac{1}{2}K^{-1}$\tabularnewline
\hline 
Lorentzian  & $(\frac{\pi K}{2})^{-\frac{1}{2}}\frac{1}{1+(z/K)^{2}}$  & $K$  & $\frac{1}{2}K^{-1}$\tabularnewline
\hline 
Exponential  & $K^{-\frac{1}{2}}\exp(-\frac{|z|}{K})$  & $\frac{1}{\sqrt{2}}K$  & $\frac{1}{\sqrt{2}}K^{-1}$\tabularnewline
\hline 
\end{tabular}\protect\protect\caption{Results for three typical examples with $\protect\ds=2$. The maximal
position shift $|\langle\Delta q\rangle|_{\max}$ can be very large
for all three examples if the spread width of the states (proportional
to $K$) is sufficiently large, or the maximal momentum shift $|\langle\Delta p\rangle|_{\max}$
can be large if $K$ is sufficiently small. This also verifies the
complementarity relation (\ref{eq:2}).}
\end{table}

\begin{figure}
\includegraphics[clip,scale=0.6]{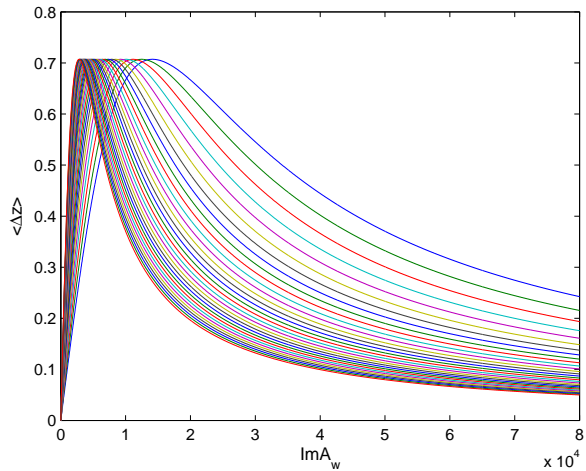}

\protect\protect\caption{(Color online) The figure shows the average shift of spin-$\frac{1}{2}$
particles $\langle\Delta z\rangle$ as $\protect\im A_{w}$ increases.
The particles are in an exponential state $\exp(-|q|)$ initially,
and the Hamiltonian is (\ref{eq:3}). Different curves are plotted
for different $-\mu\frac{\partial B_{z}}{\partial z}$, ranging from
$10^{-4}$ to $5\times10^{-4}$ (steeper curves with larger $-\mu\frac{\partial B_{z}}{\partial z}$).
Explicit turning points can be found in this figure, which indicates
the upper bound of the amplification effect. It can be seen that the
largest $\langle\Delta z\rangle$ are almost the same for different
$-\mu\frac{\partial B_{z}}{\partial z}$, which confirms the independence
from $g$ of $|\langle\Delta M\rangle|_{\max}$.}
\end{figure}

An important complementarity relation between $|\langle\Delta z\rangle|_{\max}$
and $|\langle\Delta p_{z}\rangle|_{\max}$ can be obtained from (\ref{eq:9}):
\begin{equation}
|\langle\Delta z\rangle|_{\max}|\langle\Delta p_{z}\rangle|_{\max}=\frac{1}{2}\sqrt{1+\langle\{z,\Delta p_{z}\}\rangle_{\Upsilon}^{2}}\geq\frac{1}{2}.\label{eq:2}
\end{equation}
If the wave function $\Upsilon(z)$ is real, then $\langle\{z,\Delta p_{z}\}\rangle_{\Upsilon}=0$,
so the complementarity relation becomes an equality: $|\langle\Delta z\rangle|_{\max}|\langle\Delta p_{z}\rangle|_{\max}=\frac{1}{2}$.

According to the Robertson-Schrödinger uncertainty inequality \cite{roberston uncertainty},
the second equation of (\ref{eq:9}) leads to $|\langle\Delta p_{z}\rangle|_{\max}\leq\sqrt{\langle\Delta z^{2}\rangle_{\Upsilon}}$,
so it can be combined with the second equation of (\ref{eq:9}) as
\begin{equation}
|\langle\Delta M\rangle|_{\max}\leq\sqrt{\langle\Delta M^{2}\rangle_{\Upsilon}},\, M=z,\, p_{z}.\label{eq:11}
\end{equation}

The upper bound of the amplification limit (\ref{eq:11}) has an intuitive
physical picture: the final spatial wave function of a single particle
is a superposition of several wave function that are translated very
little from the initial wave function. When the postselection on the
particle's spin is properly chosen, the superposition can cancel the
major part of the initial wave function, resulting in a large deviation
of the particle's spatial wave function from its original position.
But such a displacement cannot be too large, because the wave function
is still bounded. So the particle can be only shifted to the edge
of the spread of its wave function at most. This is what Eq.~(\ref{eq:9})
implies.

The above results can help to choose appropriate settings in designing
real weak measurement experiments. The initial detector states can
be chosen by (\ref{eq:9}) or (\ref{eq:11}) to realize a desired
amplification limit. For example, if one wants to enlarge the maximal
position shift of the detector, one should choose an initial state
with wider spread for the detector; to enlarge the maximal momentum
shift of the detector, the initial spread of the detector should be
narrower. The width of the initial detector wave function decides
the amplification limit of a weak measurement with that detector.

\subsection{Continuous variable systems}

The role of the dimension of the system is often neglected in the
study of weak measurement amplification. To show how the dimension
of the system can influence the amplification limit, we now consider
continuous variable systems, i.e., $\ds=\infty$, as a limiting case
of high dimensional systems. For simplicity, we assume $A$ and $\Omega$
are nondegenerate, i.e. $\ra=\ds$ and $\ro=\dd$.

When $\ds=\infty$, the subspace $\hd$ of all possible final detector
states can be spanned by $|\Upsilon\rangle,\Omega|\Upsilon\rangle,\cdots,\Omega^{\infty}|\Upsilon\rangle$.
Obviously, this subspace is equal to the whole Hilbert space of the
detector only if $|\Upsilon\rangle$ has full support on the eigenbasis
of $\Omega$. This implies that the final detector can be (but is
not limited to) any eigenstate of $M$. Therefore, $|\langle\Delta M\rangle|_{\max}=|\lambda(M)|_{\max}$
in this case. In particular, if the detector is also a continuous
variable system and $M=q,\, p$, then $|\langle\Delta M\rangle|_{\max}=\infty$,
but is not bounded by the spread of the detector state $\sqrt{\langle\Delta M^{2}\rangle_{\Upsilon}}$
as in (\ref{eq:11}), a dramatic difference from the results for $\ds=2$.

Of course, one cannot always postselect a continuous variable system
to be in an arbitrary state, so $|\langle\Delta M\rangle|_{\max}$
will still be finite, or only approach $\infty$ asymptotically, in
practice. Yet this limiting example demonstrates how significantly
the dimension of the system can influence the amplification limit
of a weak measurement, particularly when the system dimension is high,
which clearly shows the advantage of the variational method, since
such a result cannot be derived from the first-order weak value formalism
\cite{antonoio}. And it reveals the rich structure and complexity
of this problem.

\section{Generalization to mixed detector states}

In this section, we generalize the main results to mixed detector
states. Suppose the initial state of the detector is mixed: 
\begin{equation}
\rho_{{\rm D}}=\sum_{k}\eta_{k}|\Upsilon_{k}\rangle\langle\Upsilon_{k}|.
\end{equation}
The generalization to this case is mostly straightforward, but one
must take care to avoid a tricky pitfall. At first glance, as $\rho_{{\rm D}}$
represents the ensemble $\{\eta_{k},|\Upsilon_{k}\rangle\}$, it seems
that the maximum shift of the detector should be the average maximum
shift of the detector over the ensemble: 
\begin{equation}
|\langle\Delta M\rangle|_{\max}\stackrel{?}{=}\sum_{k}\eta_{k}|\langle\Delta M\rangle|_{\max}^{|\Upsilon_{k}\rangle},\label{eq:42}
\end{equation}
where the superscript $|\Upsilon_{k}\rangle$ indicates the dependence
of $|\langle\Delta M\rangle|_{\max}$ on $|\Upsilon_{k}\rangle$.
But the optimal choice of pre- and postselections to reach the maximum
shift depends on the initial detector state, so for different $|\Upsilon_{k}\rangle$'s,
the optimal choices of pre- and postselections are different. One
cannot make the optimal choice for all $|\Upsilon_{k}\rangle$'s simultaneously.
So (\ref{eq:42}) is not the maximal shift for a mixed detector state,
in general, but rather an upper bound on the shift.

In fact, by carrying out the previous variational procedure for mixed
detector states, it can be shown that $|\langle\Delta M\rangle|_{\max}$
is the largest absolute value over all solutions to 
\begin{equation}
\det\sum_{k}\eta_{k}(\wx{}_{k}^{\dagger}\Delta M\wx{}_{k}-\wx{}_{k}^{\dagger}\wx{}_{k}\langle\Delta M\rangle_{\e})=0,
\end{equation}
where $\wx{}_{k}$ is the matrix with $|\Upsilon_{k}\rangle,\Omega|\Upsilon_{k}\rangle,\cdots,\Omega^{\ds-1}|\Upsilon_{k}\rangle$
as its columns, as defined above. So the maximum detector shift is
\begin{equation}
\begin{aligned}|\langle\Delta M\rangle|_{\max} & =|\lambda((\sum_{i}\eta_{i}\wx_{i}^{\dagger}\wx_{i})^{-\frac{1}{2}}\sum_{k}\eta_{k}\wx_{k}^{\dagger}\Delta M\\
 & \times\wx_{k}(\sum_{j}\eta_{j}\wx_{j}^{\dagger}\wx_{j})^{-\frac{1}{2}})|_{\max},
\end{aligned}
\end{equation}
and an asymptotic solution is 
\begin{equation}
\begin{aligned} & |\langle\Delta M\rangle|_{\max}\\
= & \lim_{n\rightarrow\infty}(\tr(\sum_{k}\eta_{k}\wx_{k}^{\dagger}\Delta M\wx_{k}(\sum_{j}\eta_{j}\wx_{j}^{\dagger}\wx_{j})^{-\frac{1}{2}})^{n})^{\frac{1}{n}}.
\end{aligned}
\end{equation}

\begin{acknowledgments}
Shengshi Pang thanks Justin Dressel and Antonio Di Lorenzo for helpful
discussions. Shengshi Pang and Todd A. Brun acknowledge support from
the ARO MURI under Grant No. W911NF-11-1-0268 and the NSF grant CCF-0829870.
Shengjun Wu thanks support from the NSFC under Grant No. 11275181.
Zeng-Bing Chen thanks support from NNSF of China under Grant No. 61125502,
the CAS, the National High Technology Research and Development Program
of China, and the National Fundamental Research Program under Grant
No. 2011CB921300. 
\end{acknowledgments}

\section*{Appendix A. The subspace of the final detector states}

In this appendix, we prove by induction that $|\Upsilon\rangle,\Omega|\Upsilon\rangle,\cdots,\Omega^{(\ra-1)}|\Upsilon\rangle$
is an approximate (unorthonormalized) spanning set for the subspace
spanned by the translated detector states $\exp(-\i ga_{k}\Omega)|\Upsilon\rangle,\, k=1,\cdots,\ra$,
when $g|\lambda(A)|_{\max}\ll1$.

Suppose that $A$ has $\ra$ distinct eigenvalues. The subspace of
final detector states, which we denote as $\hd$, can be approximately
spanned by the vectors $\{\exp(-\i ga_{k}\Omega)|\Upsilon\rangle\},\, k=1,\cdots,\ra$.
Define the matrix 
\begin{equation}
\wx=\left[|\Upsilon\rangle|\,\Omega|\Upsilon\rangle\,|\,\cdots\,|\,\Omega^{\ra-1}|\Upsilon\rangle\right].
\end{equation}
It is easy to verify that the columns of $\wx(\wx^{\dagger}\wx)^{-\frac{1}{2}}$
are an orthonormal basis of $\hd$.

Now suppose that $A$ has $r_{A}+1$ distinct eigenvalues, and we
have constructed the matrix $\wx$ as above using the first $r_{A}$
eigenvalues. By the Gram-Schmidt orthogonalization procedures, the
next state in the basis (if there is one) can be obtained by 
\begin{equation}
\begin{aligned}|e_{\ra+1}\rangle & =\exp(-\i ga_{\ra+1}\Omega)|\Upsilon\rangle\\
 & -\wx(\wx^{\dagger}\wx)^{-\frac{1}{2}}(\wx^{\dagger}\wx)^{-\frac{1}{2}}\wx^{\dagger}\exp(-\i ga_{\ra+1}\Omega)|\Upsilon\rangle\\
 & =\exp(-\i ga_{\ra+1}\Omega)|\Upsilon\rangle\\
 & -\wx(\wx^{\dagger}\wx)^{-1}\wx^{\dagger}\exp(-\i ga_{\ra+1}\Omega)|\Upsilon\rangle.
\end{aligned}
\label{eq:1-1}
\end{equation}
Note that we assumed $\wx$ to have a full rank so that $\wx^{\dagger}\wx$
is invertible in (\ref{eq:1-1}). If $\wx$ does not have a full rank,
then one should use the pseudoinverse of $\wx^{\dagger}\wx$ (the
inverse on its support) instead.

Since $g|\lambda(A)|_{\max}\ll1$, 
\begin{equation}
\begin{aligned} & \exp(-\i ga_{\ra+1}\Omega)|\Upsilon\rangle\\
= & \sum_{k=0}^{\ra}\frac{(-\i ga_{\ra+1})^{k}}{k!}\Omega^{k}|\Upsilon\rangle+o(g^{\ra})\\
= & \wx X+\frac{(-\i ga_{\ra+1})^{\ra}}{\ra!}\Omega^{\ra}|\Upsilon\rangle+o(g^{\ra}),
\end{aligned}
\end{equation}
where 
\begin{equation}
X=\left(1,-\i ga_{\ra+1},\cdots,\frac{(-\i ga_{\ra+1})^{\ra-1}}{(\ra-1)!}\right)^{T}.
\end{equation}
Therefore, (\ref{eq:1-1}) can be simplified to 
\begin{equation}
\begin{aligned} & |e_{\ra+1}\rangle\\
= & \wx X+\frac{(-\i ga_{\ra+1})^{\ra}}{\ra!}\Omega^{\ra}|\Upsilon\rangle+o(g^{\ra})\\
 & -\wx(\wx^{\dagger}\wx)^{-1}\wx^{\dagger}\left(\wx X+\frac{(-\i ga_{\ra+1})^{\ra}}{\ra!}\Omega^{\ra}|\Upsilon\rangle+o(g^{\ra})\right)\\
= & \frac{(-\i ga_{\ra+1})^{\ra}}{\ra!}\left(\Omega^{\ra}|\Upsilon\rangle-\wx(\wx^{\dagger}\wx)^{-1}\wx^{\dagger}\Omega^{\ra}|\Upsilon\rangle\right)+o(g^{\ra}).
\end{aligned}
\label{eq:2-1}
\end{equation}

Since the columns of $\wx$ are $|\Upsilon\rangle,\Omega|\Upsilon\rangle,\cdots,\Omega^{\ra-1}|\Upsilon\rangle$,
it can be seen from (\ref{eq:2-1}) that $|e_{\ra+1}\rangle$ is a
linear combination of $|\Upsilon\rangle,\Omega|\Upsilon\rangle,\cdots,\Omega^{\ra}|\Upsilon\rangle$.
Thus, the subspace spanned by $\exp(-\i ga_{k}\Omega)|\Upsilon\rangle,\, k=1,\cdots,\ra+1$
can be spanned by $|\Upsilon\rangle,\Omega|\Upsilon\rangle,\cdots,\Omega^{\ra}|\Upsilon\rangle$.
This completes the proof by induction.

\section*{Appendix B. Analysis of the dimension of $\protect\hd$}

In this appendix, we analyze the dimension of the subspace $\hd$
spanned by all possible final detector states, so as to show what
determines the amplification limit $|\langle\Delta M\rangle|_{\max}$.
The vectors $|\Upsilon\rangle,\Omega|\Upsilon\rangle,\cdots,\Omega^{\ra-1}|\Upsilon\rangle$
can be chosen to be a spanning set for $\hd$, as proved in Appendix
A, so the dimension of $\hd$ is equal to the rank of the matrix $\wx=\left[|\Upsilon\rangle|\,\Omega|\Upsilon\rangle\,|\,\cdots\,|\,\Omega^{\ra-1}|\Upsilon\rangle\right]$.

Let $\Omega$ have $\ro$ distinct eigenvalues $\omega_{1},\cdots,\omega_{\ro}$,
and let the projectors of the corresponding eigensubspaces be $P_{\omega_{k}},\, k=1,\cdots,\ro$.
Then $|\Upsilon\rangle=\sum_{k=1}^{\ro}P_{\omega_{k}}|\Upsilon\rangle$,
then $\Omega^{i}|\Upsilon\rangle=\sum_{k=1}^{\ro}\omega_{k}^{i}P_{\omega_{k}}|\Upsilon\rangle$,
and $\wx=CD$, where 
\begin{equation}
C=\begin{pmatrix}c_{1}\\
 & c_{2}\\
 &  & \ddots\\
 &  &  & c_{\ro}
\end{pmatrix},\, D=\begin{pmatrix}1 & \omega_{1} & \cdots & \omega_{1}^{\ra-1}\\
1 & \omega_{2} & \cdots & \omega_{2}^{\ra-1}\\
\vdots & \vdots & \ddots & \vdots\\
1 & \omega_{\ro} & \cdots & \omega_{\ro}^{\ra-1}
\end{pmatrix}.
\end{equation}
The computational basis is $\frac{P_{\omega_{k}}|\Upsilon\rangle}{\sqrt{\langle\Upsilon|P_{\omega_{k}}|\Upsilon\rangle}}$,
and $c_{k}=\sqrt{\langle\Upsilon|P_{\omega_{k}}|\Upsilon\rangle}$.

When $|\Upsilon\rangle$ has a full support on the eigenbasis of $\Omega$,
i.e. $c_{k}\neq0$ for all $k=1,\cdots,\ro$, $\dim(\hd)=\mathrm{rank}(\wx)=\mathrm{rank}(D)$.
Since $\omega_{1},\cdots,\omega_{\ro}$ are distinct from each other,
it can be inferred that 
\begin{equation}
\dim(\hd)=\begin{cases}
\ra, & \ra\leq\ro,\\
\ro, & \ra>\ro.
\end{cases}
\end{equation}

When $|\Upsilon\rangle$ does not have full support on the eigenbasis
of $\Omega$, say $c_{n+1}=\cdots=c_{\ro}=0$, then similarly to the
above equation, it can be shown that 
\begin{equation}
\dim(\hd)=\begin{cases}
\ra, & \ra\leq n,\\
n, & \ra>n.
\end{cases}
\end{equation}

In summary, the initial detector state $|\Upsilon\rangle$, the number
of distinct eigenvalues $\ra$ and $\ro$ of $A$ and $\Omega$ (respectively),
and the support of $|\Upsilon\rangle$ in the eigenbasis of $\Omega$
determine the amplification limit.

\section*{Appendix C. General discussion for $\protect\ds=2$}

In the main text, we showed an example of spin-$\frac{1}{2}$ particles
passing through two Stern-Gerlach devices to illustrate the main results.
In this appendix, we want to give a general discussion about the results
for two dimensional systems, which is of great interest in the field
of quantum information and quantum computing. We assume $A$ to be
nondegenerate, since otherwise $A$ would be proportional to the identity
and the interaction would be trivial.

\subsection{General results for $\protect\ds=2$.}

When $\ds=2$, the subspace spanned by all possible final detector
states is approximately spanned by $|\Upsilon\rangle$ and $\Omega|\Upsilon\rangle$.
We assume that $|\Upsilon\rangle$ is not an eigenstate of $\Omega$
so that $|\Upsilon\rangle$ and $\Omega|\Upsilon\rangle$ are linearly
independent. We shall consider three typical cases below. In this
section, we shall denote $\langle\Upsilon|\cdot|\Upsilon\rangle$
by $\langle\cdot\rangle_{\Upsilon}$ for short.

For $\ds=2$, $\langle\Delta M\rangle_{\e}$ satisfies 
\begin{equation}
\begin{aligned} & \det\begin{pmatrix}-\langle\Delta M\rangle_{\e} & \langle\Delta M\Omega\rangle_{\Upsilon}-\langle\Omega\rangle_{\Upsilon}\langle\Delta M\rangle_{\e}\\
\langle\Omega\Delta M\rangle_{\Upsilon}-\langle\Omega\rangle_{\Upsilon}\langle\Delta M\rangle_{\e} & \langle\Omega\Delta M\Omega\rangle_{\Upsilon}-\langle\Omega^{2}\rangle_{\Upsilon}\langle\Delta M\rangle_{\e}
\end{pmatrix}\\
 & =0,
\end{aligned}
\end{equation}
where we have used $\langle\Delta M\rangle_{\Upsilon}=0$. $|\langle\Delta M\rangle|_{\max}$
can be straightforwardly derived from the above equation: 
\begin{equation}
|\langle\Delta M\rangle|_{\max}=\frac{W_{\Upsilon}}{2\langle\Delta\Omega^{2}\rangle_{\Upsilon}},\label{eq:6-1}
\end{equation}
where $\langle\Delta\Omega^{2}\rangle_{\Upsilon}=\langle\Omega^{2}\rangle_{\Upsilon}-\langle\Omega\rangle_{\Upsilon}^{2}$,
and 
\begin{equation}
\begin{aligned}W_{\Upsilon} & =|\langle\Omega\rangle_{\Upsilon}\langle\{\Omega,\Delta M\}\rangle_{\Upsilon}-\langle\Omega\Delta M\Omega\rangle_{\Upsilon}|\\
 & +((\langle\Omega\rangle_{\Upsilon}\langle\{\Omega,\Delta M\}\rangle_{\Upsilon}-\langle\Omega\Delta M\Omega\rangle_{\Upsilon})^{2}\\
 & +\langle\Delta\Omega^{2}\rangle_{\Upsilon}(\langle\{\Omega,\Delta M\}\rangle_{\Upsilon}^{2}-\langle[\Omega,\Delta M]\rangle_{\Upsilon}^{2}))^{\frac{1}{2}}.
\end{aligned}
\end{equation}

Note that $\langle\Delta M\rangle_{\Upsilon}=0$, so according to
the Robertson-Schrödinger uncertainty inequality \cite{roberston uncertainty},
an upper bound for $|\langle\Delta M\rangle|_{\max}$ is 
\begin{equation}
\begin{aligned}|\langle\Delta M\rangle|_{\max} & \leq\frac{|\langle\Omega\rangle_{\Upsilon}\langle\{\Omega,\Delta M\}\rangle_{\Upsilon}-\langle\Omega\Delta M\Omega\rangle_{\Upsilon}|}{\langle\Delta\Omega^{2}\rangle_{\Upsilon}}\\
 & +\sqrt{\langle\Delta M^{2}\rangle_{\Upsilon}},
\end{aligned}
\label{eq:3-2}
\end{equation}
where $\langle\Delta M^{2}\rangle_{\Upsilon}=\langle M^{2}\rangle_{\Upsilon}-\langle M\rangle_{\Upsilon}^{2}$.

\subsection{Symmetric detector states}

Now, we assume that the initial detector state $|\Upsilon\rangle$
is symmetric about its center, and that $\Omega=q$, $M=q,\, p$.
In this case, $\langle\Omega\rangle_{\Upsilon}=\langle\Omega\Delta M\Omega\rangle_{\Upsilon}=0$,
so (\ref{eq:3-2}) implies that the maximal shift of the detector
will not exceed the spread of the initial detector states in the eigenbasis
of $M$. According to (\ref{eq:6-1}) the exact maximal shift of the
detector should be 
\begin{equation}
|\langle\Delta q\rangle|_{\max}=\sqrt{\langle\Delta q^{2}\rangle_{\Upsilon}},\ |\langle\Delta p\rangle|_{\max}=\frac{1}{2}\sqrt{\frac{1+\langle\{q,\Delta p\}\rangle_{\Upsilon}^{2}}{\langle\Delta q^{2}\rangle_{\Upsilon}}}.\label{eq:9-1}
\end{equation}

An important property of $|\langle\Delta q\rangle|_{\max}$ and $|\langle\Delta p\rangle|_{\max}$
is that 
\begin{equation}
|\langle\Delta q\rangle|_{\max}|\langle\Delta p\rangle|_{\max}=\frac{1}{2}\sqrt{1+\langle\{q,\Delta p\}\rangle_{\Upsilon}^{2}}\geq\frac{1}{2},\label{eq:10}
\end{equation}
which indicates a complementarity relationship between $|\langle\Delta q\rangle|_{\max}$
and $|\langle\Delta p\rangle|_{\max}$. A special case is that when
the wave function of the initial detector state is real, then $\langle\{q,\Delta p\}\rangle_{\Upsilon}=0$,
and (\ref{eq:10}) becomes an equality: 
\begin{equation}
|\langle\Delta q\rangle|_{\max}|\langle\Delta p\rangle|_{\max}=\frac{1}{2}.
\end{equation}

\end{document}